\documentclass[aps,prb,twocolumn,showpacs,floatfix,superscriptaddress,nofootinbib]{revtex4}
\usepackage{graphicx}
\usepackage{hyperref}
\usepackage{amssymb}
\usepackage{amsmath}
\usepackage{lmodern}
\usepackage{color}
\usepackage{empheq}
\usepackage{epstopdf}
\usepackage{amsmath}

\DeclareMathOperator{\sech}{sech}

\begin{document}

\title{Shock waves in  nonlinear transmission lines}

\author{Eugene Kogan}
\email{Eugene.Kogan@biu.ac.il}
\affiliation{Department of Physics, Bar-Ilan University, Ramat-Gan 52900, Israel}

\begin{abstract}

In the first half of the paper we consider interaction between the small amplitude travelling waves ("sound") and the shock waves in the   transmission line
containing both nonlinear capacitors and nonlinear inductors.   We calculate
 the "sound" wave  coefficient of reflection   from (coefficient of transmission through)  the shock wave. These coefficients are expressed in terms of the speeds of the "sound" waves relative to the shock and the wave impedances. In the second half of the paper
we explicitly include into consideration the dissipation in the system, introducing ohmic resistors shunting the inductors and also in series with the capacitors. This allows us to justify the conditions on the shocks, postulated in the first half of the paper. This also allows us to
describe the shocks as physical objects of finite width and study their profiles,
same as the profiles of the waves closely connected with the shocks - the kinks. The profiles of the latter, and 
in some particular cases  the profiles of the former, were obtained in terms of elementary functions.

\end{abstract}

\date{\today}

\maketitle

\section{Introduction}

The  nonlinear electrical transmission lines are of much interest both due to their applications, and as the laboratories to study nonlinear waves \cite{malomed2}.
A very interesting particular type of signals which can propagate along such
lines - the shock waves - is attracting interest since long ago \cite{landauer1,landauer2}.
We published a series of papers on the travelling waves in Josephson transmission lines (JTL): the kinks, the solitons and the shocks \cite{kogan4} (see the last publication of that series and the references therein). The cited publication is especially relevant for what will be presented below. In that publication we considered among other problems, the small amplitude wave ("sound") reflection from and transmission through the shock wave.

Note that from the point of view of an electrical engineer, Josephson junctions  are just the  nonlinear inductors, and this is how  they were described  in all our papers on JTL. On the other hand, the capacitors in all these papers were assumed to be linear. On the other hand,
very recently we published the paper on the travelling waves in the  transmission
line with the linear inductors but with the nonlinear capacitors \cite{kogan5}.

In the present paper we consider  the travelling waves in the transmission line where, in the general case, both the inductors and the capacitors are nonlinear.
In the first part of the paper, generalizing our previous results \cite{kogan4,kogan5}, we calculate the "sound"
reflection from and  transmission through the shock wave in these  nonlinear transmission lines. In the second part of the paper we include explicitly the ohmic losses (which are of course necessary for the existence of the shock waves) into the circuit equations.

\section{The circuit equations}
\label{circ}

\begin{figure}[h]
\includegraphics[width=.9\columnwidth]{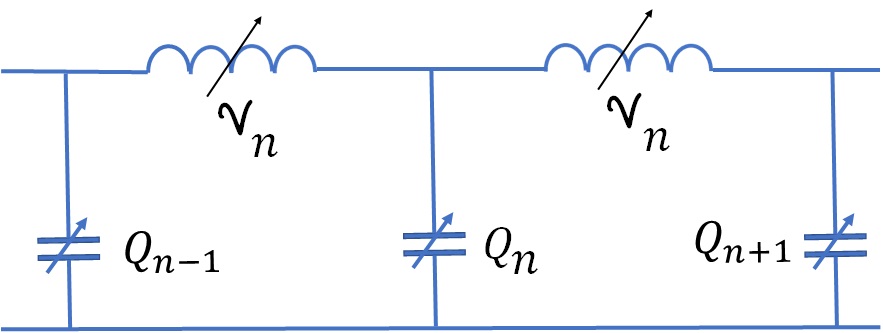}
\caption{Lossless nonlinear  transmission line}
 \label{trans5}
\end{figure}

The   transmission line constructed from the identical nonlinear inductors and the identical nonlinear capacitors is shown on Fig. \ref{trans5}.
We take the  capacitors charges $Q_n$ and the integrated voltages on the inductors
\begin{eqnarray}
\Phi_n\equiv\int {\cal V}_ndt
\end{eqnarray}
as the dynamical variables.
The circuit equations (Kirchhoff laws) are
\begin{subequations}
\label{a10}
\begin{alignat}{4}
\frac{dQ_n}{dt}&=  I_n-I_{n+1},\label{a8b}\\
\frac{d \Phi_{n+1}}{d t}&=V_{n}-V_{n+1}.\label{8a}
\end{alignat}
\end{subequations}
To close the system (\ref{a10}) we should specify the connection between the voltages on the capacitors $V_n$ and the charges and also between the currents through the inductors $I_n$ and the integrated voltages
\begin{subequations}
\label{q}
\begin{alignat}{4}
V_n&= V(Q_n),\label{qq}\\
I_n&=I(\Phi_n).\label{qq2}
\end{alignat}
\end{subequations}
Further on we'll consider $V(Q)$ and $I(\Phi)$ as known functions.

In the continuum approximation we  treat $n$  as the continuous variable $x$
(we measure distance in the units of the transmission line period)
and approximate the finite differences in the r.h.s. of the equations by the first derivatives with respect to $x$,   after which the equations take the form
\begin{subequations}
\label{ve9c}
\begin{alignat}{4}
\frac{\partial Q}{\partial t} &=  -\frac{\partial I}{\partial x},\label{vb}\\
\frac{\partial \Phi}{\partial t}&= -\frac{\partial V}{\partial x}. \label{vb0}
\end{alignat}
\end{subequations}

\section{The "sound" and the shocks}
\label{sou}

Because  (\ref{ve9c})  is nonlinear, it can have a lot of different types of solutions.
In this paper we'll be interested in only two types of those. First type -
small amplitude  waves
on a homogeneous background $ Q_0,\Phi_0$.
\begin{subequations}
\begin{alignat}{4}
Q&= Q_0+q,\\
\Phi&=\Phi_0+\phi.
\end{alignat}
\end{subequations}
For such waves Eq. (\ref{ve9c}) is simplified to
\begin{subequations}
\label{ve9d}
\begin{alignat}{4}
\frac{\partial q}{\partial t} &=  -\frac{1}{L_0}\frac{\partial \phi}{\partial x},\\
\frac{\partial \phi}{\partial t}&= -\frac{1}{C_0}\frac{\partial q}{\partial x},
\end{alignat}
\end{subequations}
where
\begin{eqnarray}
\frac{1}{C_0}= \left.\frac{dV}{dQ}\right|_{Q=Q_0},\hskip 1cm
\frac{1}{L_0}= \left.\frac{dI}{d\Phi}\right|_{\Phi=\Phi_0}.
\end{eqnarray}

The  solutions of Eq. (\ref{ve9d}) (which, for brevity we'll call the "sound") are right- and left-propagating travelling waves, each   depending upon the single variable
$\tau_{\pm}=t\mp x/u_0$. The variables in the "sound" wave are
 connected by the equations
\begin{subequations}
\begin{alignat}{4}
v&=\pm u_0\phi,\label{1}\\
i&=\pm u_0 q,\label{2}\\
\phi &=\pm Z_0q,\label{3}
\end{alignat}
\end{subequations}
where
\begin{subequations}
\begin{alignat}{4}
 u_0&=\frac{1}{\sqrt{L_0 C_0}}, \label{u0}\\
Z_0&=\sqrt{\frac{L_0}{C_0}}\label{z0}
\end{alignat}
\end{subequations}
are the "sound" speed and
 the linear impedance of the transmission line respectively.

The second type of solutions, we are  interested in, is shock waves, containing (in the approximation of the present Section) discontinuities, satisfying the conditions (see Section \ref{lossy})
\begin{subequations}
\label{3ve9c}
\begin{alignat}{4}
U_{21}\left(Q_2-Q_1\right) &=  I_2-I_1, \label{3vb}\\
U_{21}\left(\Phi_2-\Phi_1\right)&= V_2-V_1, \label{3vb0}
\end{alignat}
\end{subequations}
where
$V_{a,b}\equiv V(Q_{a,b})$,  $I_{a,b}\equiv I(\Phi_{a,b})$,
and  $U_{21}$ is the  shock wave speed.
The indices 1 and 2, which  refer to the quantities before and after the shock respectively,  enter  in the system
(\ref{3ve9c}) in a symmetrical way.  However, in Section \ref{lossy} we'll show
 that the shock always propagates in such a way that \cite{whitham,kogan4,kogan5}
\begin{eqnarray}
\label{kivun}
u_2^2>U_{21}^2>u_1^2.
\end{eqnarray}
The shock is supersonic with respect to the phase before the front, but subsonic with respect to the phase behind the front.

\section{The "sound" reflection and transmission}
\label{sou2}

\begin{figure}[h]
\includegraphics[width=.6\columnwidth]{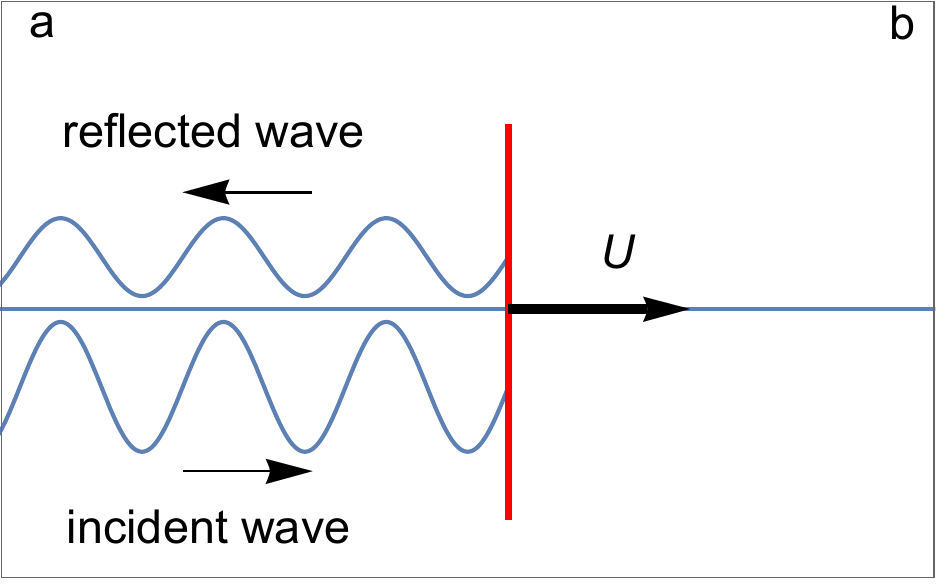}
\caption{Reflection of a "sound" wave from a shock wave.}
\label{reflection}
\end{figure}

\begin{figure}[h]
\includegraphics[width=.6\columnwidth]{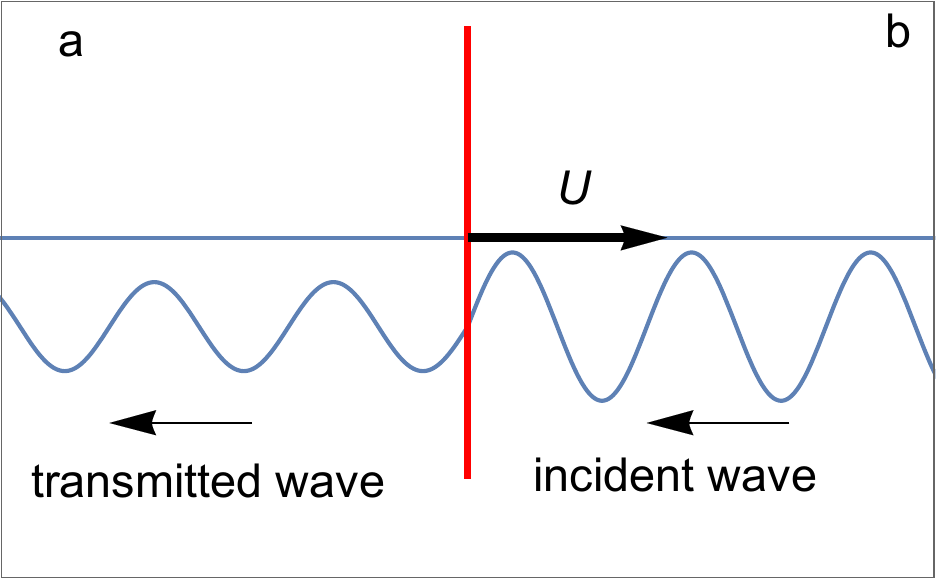}
\caption{Transmission of a "sound" wave through a shock wave.}
\label{transmission}
\end{figure}

We'll be interested in two problems \cite{landau,congy}.
The first one: A  "sound" wave is incident  from the rear  on a shock wave (see Fig.
\ref{reflection}).
The second problem: A  "sound" wave is incident  from the front on a shock wave (see Fig. \ref{transmission}).
(Note the difference between the problems we consider and the problem of  propagation of high-frequency wave packets along a smooth evolving background flow \cite{kamchatnov}.)
While formulating both problems we took into account the inequalities (\ref{kivun}),
meaning that there is no "sound" transmission in the first case, and no "sound" reflection in the second case.

For the first problem mentioned above we have
\begin{eqnarray}
\label{in}
\left(\begin{array}{c|c}Q_1 &Q_2\\\Phi_1 & \Phi_2\end{array}\right)
=\left(\begin{array}{c|c}Q_b &Q_a+q^{(in)}+ q^{(r)}\\\Phi_b &
\Phi_a+\phi^{(in)}+ \phi^{(r)}\end{array}\right),
\end{eqnarray}
where
(in) stands for the incident  and (r) - for the reflected "sound" wave.
 We also have to take into account the change of the shock speed due to the "sound" waves
\begin{eqnarray}
U_{21}=U_{ab}+\delta U
\end{eqnarray}
(same in the next problem).
Substituting (\ref{in}) into (\ref{3ve9c}),
 in linear with respect to the "sound" wave amplitudes approximation we obtain
 (taking into account (\ref{1}) and (\ref{2}))
\begin{subequations}
\label{in2}
\begin{alignat}{4}
\delta U\left(Q_a-Q_b\right)
+U_{ab}\left( q^{(r)}+q^{(in)}\right)
=u_a\left(q^{(in)}-q^{(r)}\right),\\
\delta U\left(\Phi_a-\Phi_b\right)+ U_{ab}\left(\phi^{(r)}+\phi^{(in)}\right)
=u_a\left(\phi^{(in)}-\phi^{(r)}\right).
\end{alignat}
\end{subequations}
Excluding $\delta U$ from  (\ref{in2})  and taking into account (\ref{3}) we obtain for the reflection coefficient
\begin{eqnarray}
\label{rr0}
R\equiv \frac{ \phi^{(r)}}{\phi^{(in)}}
=\frac{\left(Z_a-Z_{ab}\right)u_{in}}
{\left(Z_a+Z_{ab}\right)u_{r}},
\end{eqnarray}
where $u_{in}=u_a-U_{ab}$ is the speed of the incident "sound" wave relative to the shock wave, $u_{r}=u_a+U_{ab}$ is the speed of the reflected "sound" wave relative to the shock wave and we introduced the impedance of the shock wave
\begin{eqnarray}
\label{zzz}
Z_{ab}\equiv\frac{\Phi_a-\Phi_b}{Q_a-Q_b}.
\end{eqnarray}
Note that from (\ref{3ve9c}) follows
\begin{eqnarray}
\label{cons}
\frac{\Phi_a-\Phi_b}{Q_a-Q_b}=\frac{V_a-V_b}{I_a-I_b}.
\end{eqnarray}

For the second problem we have
\begin{eqnarray}
\label{out}
\left(\begin{array}{c|c}Q_1 &Q_2\\\Phi_1 & \Phi_2\end{array}\right)
=\left(\begin{array}{c|c}Q_b+q^{(in)}&Q_a+ q^{(t)}\\
\Phi_b+\phi^{(in)} &\Phi_a+\phi^{(t)}\end{array}\right),
\end{eqnarray}
where (t) stands for the transmitted wave.
Substituting  (\ref{out})  into (\ref{3ve9c}),
in linear approximation we obtain
\begin{subequations}
\label{out2}
\begin{alignat}{4}
\delta U\left(Q_a-Q_b\right)
+U_{ab}\left(q^{(t)}-q^{(in)}\right)=u_bq^{(in)}-u_aq^{(t)},\\
\delta U\left(\Phi_a-\Phi_b\right)+ U_{ab}\left(\phi^{(t)}-\phi^{(in)}\right)
=u_b\phi^{(in)}-u_a\phi^{(t)}.
\end{alignat}
\end{subequations}
Excluding $\delta U$ from  (\ref{out2}) and taking into account (\ref{3}) we obtain for the transmission coefficient
\begin{eqnarray}
\label{rr9}
T\equiv \frac{ \phi^{(t)}}{ \phi^{(in)}}
=\frac{\left(Z_b+Z_{ab}\right)Z_au_{in}}
{\left(Z_a+Z_{ab}\right)Z_bu_{t}},
\end{eqnarray}
where $u_{in}=u_b
+U_{ab}$ is the speed of the incident "sound" wave relative to the shock wave, and $u_{t}=u_a+U_{ab}$ is the speed of the transmitted "sound" wave relative to the shock wave.

As one could have expected, the modulus of the "sound" reflection coefficient is less than one, and it goes to zero when the intensity of the shock wave decreases, that is when $Q_a\to Q_b,\Phi_a\to \Phi_b$,
in other words, when the shock wave itself is nearly  the "sound" wave.
Similarly,  in these circumstances the transmission coefficient goes to 1.
Looking  at  (\ref{rr0}) and (\ref{rr9})
we wonder whether these equations could have been obtained  by considering propagation of the waves in the frame of reference moving with the shock wave.

\subsection{Half-nonlinear transmission line}

\subsubsection{Linear capacitor}

For  linear capacitor ($V=Q/C$, where $C$ is a constant)
\begin{eqnarray}
Z_{a,b}=\frac{1}{Cu_{a,b}},\hskip 1cm
Z_{ab}=\frac{1}{CU_{ab}}.
\end{eqnarray}
Hence Eqs. (\ref{rr0}) and (\ref{rr9}) can be simplified respectively to
\begin{subequations}
\label{r0}
\begin{alignat}{4}
R&=-\frac{u_{in}^2}{u_{r}^2},\\
T&=\frac{u_{in}^2}{u_{t}^2},
\end{alignat}
\end{subequations}
and from (\ref{u0}) and (\ref{3ve9c}) follows
\begin{subequations}
\label{uuu}
\begin{alignat}{4}
u_{a,b}^2&=\frac{1}{C}\left.\frac{dI}{d\Phi}\right|_{\Phi=\Phi_{a,b}} \label{uuua}\\
U_{ab}^2&=\frac{1}{C}\frac{I_a-I_b}{\Phi_a-\Phi_b} \label{uuub}.
\end{alignat}
\end{subequations}
For example, for the JTL
\begin{eqnarray}
I=I_c\sin\left(\frac{2e\Phi}{\hbar}\right),
\end{eqnarray}
and all the velocities can be easily calculated.
Equations (\ref{r0})   in this case were
obtained in our previous publication \cite{kogan4}.

\subsubsection{Linear inductor}

For  linear inductor ($I=\Phi/L$, where $L$ is a constant)
\begin{eqnarray}
Z_{a,b}=Lu_{a,b},\hskip 1cm
Z_{ab}=LU_{ab}.
\end{eqnarray}
Hence Eqs. (\ref{rr0}) and (\ref{rr9})
can be respectively simplified to
\begin{subequations}
\label{rt}
\begin{alignat}{4}
R&=\frac{u_{in}^2}{u_{r}^2},\\
T&=\frac{Z_au_{in}^2}{Z_bu_{t}^2},
\end{alignat}
\end{subequations}
and
from (\ref{u0}) and (\ref{3ve9c})  follows
\begin{subequations}
\label{u}
\begin{alignat}{4}
u_{a,b}^2&=\frac{1}{L}\left.\frac{dV}{dQ}\right|_{Q=Q_{a,b}} \label{ua}\\
U_{ab}^2&=\frac{1}{L}\frac{V_a-V_b}{Q_a-Q_b} \label{ub}.
\end{alignat}
\end{subequations}

\section{Lossy transmission line}
\label{lossy}

As it is well known \cite{landau}, the shock waves exist only in the presence of dissipation. (In the absence of the dissipation, the localized travelling waves in the systems we consider are kinks and solitons \cite{kogan1,kogan2,kogan4}.) On the other hand, many properties of the shocks in fluids were studied beginning from the XIX century without taking the dissipation explicitly into account, but just postulating the existence of discontinuous solutions and the jump conditions \cite{krehl}.
This is exactly what we did in the previous Sections.

\begin{figure}[h]
\includegraphics[width=\columnwidth]{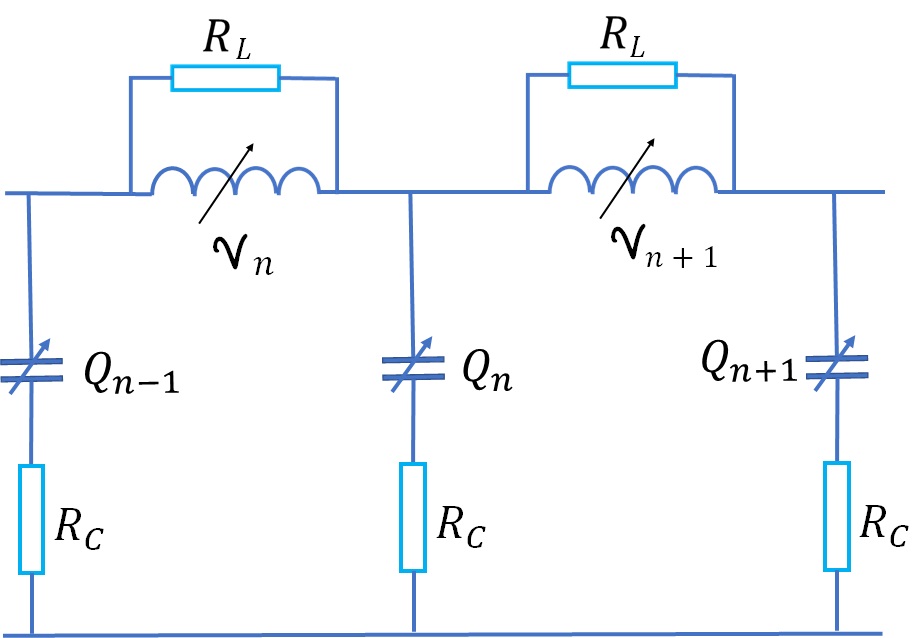}
\caption{Lossy nonlinear  transmission line}
 \label{r5}
\end{figure}

 However in this Section,  to be consistent, we  assume the existence of ohmic resistors in the transmission line, shunting the inductors and  in series with the capacitors. Such transmission line is shown on Fig. \ref{r5}. We'll resolve the zero width discontinuities from the previous Sections, which will turn out to be narrow (for weak dissipation) crossover regions and rigorously prove the jump conditions.

The corrected Eqs. (\ref{a10}) are
\begin{subequations}
\label{l}
\begin{alignat}{4}
\frac{dQ_n}{dt}&=  I_n-I_{n+1}+\frac{1}{R_L}\frac{d}{dt}\left(\Phi_n-\Phi_{n+1}\right),\label{lb}\\
\frac{d \Phi_{n+1}}{d t}&
=V_{n}-V_{n+1}+R_C\frac{d}{dt}\left(Q_n-Q_{n+1}\right).\label{la}
\end{alignat}
\end{subequations}
In the continuum approximation (\ref{l}) takes the form (compare with (\ref{ve9c}))
\begin{subequations}
\label{c2}
\begin{alignat}{4}
\frac{\partial Q}{\partial t} &=  -\frac{\partial I}{\partial x}
 -\frac{1}{R_L}\frac{\partial^2 \Phi}{\partial x\partial t},\\
\frac{\partial \Phi}{\partial t}&= -\frac{\partial V}{\partial x}
-R_C\frac{\partial^2 Q}{\partial x\partial t}.
\end{alignat}
\end{subequations}
Further on we'll consider  the  travelling waves, for which all the dependent variables   depend upon the single independent variable $\tau=t-x/U$, where $U$ is the speed of the wave. For such waves (\ref{c2}) turns into the system of ODE
which   after integration takes the form
\begin{subequations}
\label{e}
\begin{alignat}{4}
\frac{1}{R_L}\frac{d \Phi}{d\tau}&=UQ-I(\Phi)+\text{const}_1,\label{eb}\\
R_C\frac{d Q}{d\tau}&=U\Phi-V(Q)+\text{const}_2, \label{ea}
\end{alignat}
\end{subequations}

Equation (\ref{e}) contains 3 arbitrary constants - $U,\text{const}_1,\text{const}_2$
- hence to define the solution unequivocally we need 5 boundary conditions. Because we are ready to consider the solution up to translation  in time, 4 of those would be enough. 
Since we are considering localized travelling waves, 
it would be convenient to take as those the quantities $Q_1,Q_2,\Phi_1,\Phi_2$, defined by the equations
\begin{subequations}
\label{bc}
\begin{alignat}{4}
\lim_{\tau\to-\infty}Q(\tau)&=Q_1; \hskip .5cm \lim_{\tau\to+\infty}Q(\tau)=Q_2; \label{bca}\\
\lim_{\tau\to-\infty}\Phi(\tau)&=\Phi_1; \hskip .5cm \lim_{\tau\to+\infty}\Phi(\tau)=\Phi_2.
\label{bcb}
\end{alignat}
\end{subequations}
Equating the r.h.s. of (\ref{eb}) at $\tau=\pm\infty$,  we recover the jump conditions (\ref{3vb}). Doing the same  with (\ref{ea}) we recover the jump conditions (\ref{3vb0}).

The boundary conditions show that $(Q_1,\Phi_1)$ is the semistable fixed point of the system (\ref{e}), and
$(Q_2,\Phi_2)$ is the stable fixed point (see, however, Section \ref{vs}).  (Other options would have demanded fine tuning of the parameters.) This means
\begin{eqnarray}
\left|\begin{array}{cc}
\frac{d}{d\Phi}\left[U_{21}Q-I(\Phi)\right]_{Q_2,\Phi_2} &
\frac{d}{dQ}\left[U_{21}Q-I(\Phi)\right]_{Q_2,\Phi_2} \\
\frac{d}{d\Phi}\left[U_{21}\Phi-V(Q)\right]_{Q_2,\Phi_2} &\frac{d}{dQ}\left[U_{21}\Phi-V(Q)\right]_{Q_2,\Phi_2}
\end{array}\right|>0,\nonumber\\
\end{eqnarray}
or
\begin{eqnarray}
\label{z}
\left|\begin{array}{cc}
 -\left.\frac{dI}{d\Phi}\right|_{\Phi_2} & U_{21}  \\
 U_{21} & -\left.\frac{dV}{dQ}\right|_{Q_2}
\end{array}\right|>0.
\end{eqnarray}
Calculating the determinant and recalling the result  for the "sound" speed we obtain
\begin{eqnarray}
\label{ma}
U_{21}^2<u_2^2.
\end{eqnarray}
Similarly for
the point $(Q_1,\Phi_1)$  we obtain
\begin{eqnarray}
\label{mi}
u_1^2<U_{21}^2.
\end{eqnarray}
Thus we proved the inequalities (\ref{kivun}) postulated previously.

\section{Half-nonlinear transmission line}
\label{hf}

Further analysis of  the fully nonlinear line we relegate to Appendix \ref{full})
and here consider transmission line with the linear capacitor.
Because of the  symmetry of Eq. (\ref{e})
the case of the transmission line with  linear capacitors and nonlinear inductors (JTL first and foremost) would be identical to the one considered in this Section.
We can exclude the current from (\ref{e})  and after a bit of algebra obtain
closed equation for $Q(\tau)$
\begin{eqnarray}
\label{199}
\frac{LR_C}{R_L}\frac{d^2Q}{d\tau^2}+R_C\frac{dQ}{d\tau}
+\frac{L}{R_L}\frac{dV(Q)}{d\tau}=F(Q),
\end{eqnarray}
where
\begin{eqnarray}
\label{f}
F(Q)\equiv\frac{V_2-V_1}{Q_2-Q_1}Q-V(Q)+\frac{V_1Q_2-V_2 Q_1}{Q_2-Q_1}.
\end{eqnarray}
The solution of (\ref{199}) should satisfy the boundary conditions (\ref{bca}).
(The third  constant appears when we connect $\Phi$ with the found $Q$.)

Equation (\ref{199}) describes motion with friction of a fictitious Newtonian particle, $Q$ being the coordinate of the particle, in the potential well $\Pi(Q)$ defined by the equation
\begin{eqnarray}
-\frac{d\Pi(Q)}{dQ}=F(Q).
\end{eqnarray}
The boundary conditions (\ref{bc}) tell that the fictitious particle starts its motion at $\tau=-\infty$ at the potential maximum of $\Pi(Q)$ at $Q=Q_1$ and ends - at $\tau=+\infty$ at the  potential minimum at $Q=Q_2$  (see, however, Section \ref{vs}).
We may present
these maximum/minimum conditions as
\begin{eqnarray}
\label{52}
\left.\frac{dF(Q)}{dQ}\right|_{Q=Q_2}
=\frac{V_2-V_1}{Q_2-Q_1}-\left.\frac{dV}{dQ}\right|_{Q=Q_2}<0
\end{eqnarray}
and opposite inequality for $V=V_1$. Note that 
we have reproduced (\ref{z}) in the particular case of half-nonlinear transmission line.
It would be convenient further on to rewrite (\ref{f}) as
\begin{eqnarray}
\label{pp2}
F(Q)=\frac{\left[V(Q)-V_1\right]\left(Q-Q_2\right)-\left\{2\leftrightarrows 1\right\}}{Q_2-Q_1},
\end{eqnarray}
where the notation $\left\{2\leftrightarrows 1\right\}$ means the preceding  term with the indexes $2$ and $1$ interchanged.

Equation (\ref{199}) can be integrated numerically, however in this paper we would be looking for analytical solutions.
Trivially, for weak shocks, when $|Q_1-Q_2|$ is small, 
we can discard the term with the second derivative in the l.h.s. of the equation and it can be integrated in quadratures. However we'll see that (\ref{199}) can be integrated analytically (and even in terms of elementary functions) also in the case when both terms in the l.h.s. of the equation are of the same order of magnitude, though under additional assumptions.
 
Let us   approximate  $V(Q)$  for $Q$ between $Q_1$  and $Q_2$  as
\begin{eqnarray}
\label{we}
V(Q)=\text{const}+\frac{Q+\beta Q^2}{C_s},
\end{eqnarray}
where $\beta$ and $C_s$ are constants.
Let us also take $V(Q)$ in the last term in the l.h.s. of (\ref{199}) in the linear approximation.
In this cases,   (\ref{199}) becomes
\begin{eqnarray}
\label{29}
T^2\frac{d^2Q}{d\tau^2}+\tau_R\frac{dQ}{d\tau}
=\beta(Q-Q_1)(Q-Q_2),
\end{eqnarray}
where
\begin{subequations}
\begin{alignat}{4}
T^2&\equiv \frac{LC_sR_C}{R_L},\\
\tau_R&\equiv R_CC_s+\frac{L}{R_L}.
\end{alignat}
\end{subequations}
Note that
\begin{eqnarray}
\Pi(Q_1)-\Pi(Q_2)=\frac{\beta}{6}\left(Q_1-Q_2\right)\left(Q_1^2-Q_1Q_2+Q_2^2\right),
\nonumber\\
\end{eqnarray}
hence
\begin{eqnarray}
\beta(Q_1-Q_2)>0.
\end{eqnarray}

We can also improve the approximation (\ref{we}) by modifying it to
\begin{eqnarray}
\label{cube}
V(Q)=\text{const}+\frac{Q+\beta Q^2+\gamma Q^3}{C_s},
\end{eqnarray}
 where $\gamma$ is also  assumed to be constant.
This modification can be important when  $Q_1,Q_2$ are close to zero and, as it normally happens, the charge is an odd function of the voltage. 
In this cases,   (\ref{199}) becomes
\begin{eqnarray}
\label{29b}
T^2\frac{d^2Q}{d\tau^2}+\tau_R\frac{dQ}{d\tau}
=\gamma(Q-Q_1)(Q-Q_2)(Q+Q_3),
\end{eqnarray}
where
\begin{eqnarray}
Q_3\equiv \frac{\beta}{\gamma}+Q_1+Q_2.
\end{eqnarray}

\section{The ODE which doesn't contain explicitly the independent variable}
\label{re}

Following the well known in mathematics principle, stating that the more general the problem is, the easier it is to solve it, let us change gears and instead of (\ref{29b}) consider
the generalized  damped Helmholtz-Duffing  equation \cite{kogan4}
\begin{eqnarray}
\label{99}
x_{\tau\tau}+k x_{\tau}
=\gamma x\left(x^n-x_1\right)\left(x^n+x_3\right),
\end{eqnarray}
where $n$, $k$, $\gamma$,  $x_1$, $x_3$  are  constants,
with the boundary conditions
\begin{eqnarray}
\label{b}
\lim_{\tau\to-\infty}x(\tau)=x_1^{1/n},\hskip 1 cm \lim_{\tau\to+\infty}x(\tau)=0.
\end{eqnarray}

Equation (\ref{99})  doesn't contain explicitly the independent variable $\tau$. This prompts the idea to
consider $x$ as the new independent variable and
\begin{eqnarray}
\label{cot}
p=\frac{dx}{d\tau}
\end{eqnarray}
as the new dependent variable.
In the new variables
(\ref{99})  takes the form  of Abel equation of the second kind   \cite{polyanin}.
\begin{eqnarray}
\label{pp}
pp_{x}+k p=\gamma x\left(x^n-x_1\right)\left(x^n+x_3\right).
\end{eqnarray}
The boundary conditions in the new variables are
\begin{eqnarray}
\label{bmp}
p\left(x_1^{1/n}\right)= p(0)=0.
\end{eqnarray}
One can easily check up that for
$\gamma$ and $k$ connected by the formula
\begin{eqnarray}
\label{kkk}
k^2=\frac{\gamma\left[x_1+(n+1)x_3\right]^2}{n+1},
\end{eqnarray}
the solution of (\ref{pp}) satisfying the boundary conditions (\ref{bmp}) is
\begin{eqnarray}
\label{pmx}
p=mx\left(x^n-x_1\right),\hskip 1cm
m=\sqrt{\frac{\gamma}{n+1} }.
\end{eqnarray}
Substituting $p(x)$ into (\ref{cot})  and integrating  we obtain the solution of (\ref{99}) as
\begin{eqnarray}
\label{mm}
x(\tau)=\frac{x_1^{1/n}.}
{\left[\exp\left(nmx_1\tau\right)+1\right]^{1/n}}.
\end{eqnarray}

Now consider the equation
\begin{eqnarray}
\label{19}
x_{\tau\tau}+k x_{\tau}
=\beta x\left(x^n-x_1\right)
\end{eqnarray}
with the boundary conditions (\ref{b}).
We can present (\ref{19}) as
\begin{eqnarray}
\label{19b}
x_{\tau\tau}+k x_{\tau}
=\beta x\left(x^{n/2}-x_1^{1/2}\right)\left(x^{n/2}+x_1^{1/2}\right).
\end{eqnarray}
Hence we realize that for
$k$ and $\beta$ connected by the formula
\begin{eqnarray}
k=(n+4)\sqrt{\frac{\beta x_1}{2(n+2)}},
\end{eqnarray}
the solution of (\ref{19}) satisfying the boundary conditions (\ref{bmp}) is
\begin{eqnarray}
\label{mm2}
x(\tau)=\frac{x_1^{1/n}.}
{\left\{\exp\left[n\sqrt{\frac{\beta x_1}{2(n+2)}}\tau\right]+1\right\}^{2/n}}.
\end{eqnarray}

To additionally illustrate the method of integration used in this Section, in the Appendix \ref{dra}
we recover in the framework of the method the solitary wave solution of a generalized KdV equation \cite{drazin}

Let us return to (\ref{99})  and modify it to
\begin{eqnarray}
\label{99b}
x_{\tau\tau}+k(1+2\beta x^n) x_{\tau}
=\gamma x\left(x^n-x_1\right)\left(x^n+x_3\right).
\end{eqnarray}
Thus we take into account possible nonlinearity of friction. Thus instead of (\ref{pp}) we obtain
\begin{eqnarray}
\label{pp2b}
pp_{x}+k(1+2\beta x^n) p=\gamma x\left(x^n-x_1\right)\left(x^n+x_3\right).
\end{eqnarray}
Acting as above we obtain
that for
$k$ and $\gamma$ connected by the formula
\begin{eqnarray}
\label{kk}
k^2=\frac{\gamma [x_1+(n+1)x_3]^2}{\left(1-2\beta x_3\right)
\left(n+1+2\beta x_1\right)},
\end{eqnarray}
the solution of (\ref{pp2b}) satisfying the boundary conditions (\ref{bmp}) is
(\ref{pmx}) (same as it was for $\beta=0$), only
 this time
\begin{eqnarray}
\label{mk}
m= \sqrt{\frac{\gamma\left(1-2\beta x_3\right)}{n+1+2\beta x_1}}.
\end{eqnarray}
Hence  the solution of (\ref{99b}) with $\beta\neq 0$
is of the same form   as for $\beta=0$ (Eq. (\ref{mm})), only with the modified $m$.

\section{Back to the transmission line}

Now let us return to the transmission line.
For (\ref{29b}),  using the results of the previous Section, we claim that when the parameters of the equation  satisfy the relation
 \begin{eqnarray}
\label{x}
\frac{\tau_R^2}{T^2}=\frac{\gamma\left(Q_1+Q_2+2Q_3\right)^2}{2},
\end{eqnarray}
the solution of  the equation  can be expressed in terms of elementary functions:
\begin{eqnarray}
\label{oib}
Q=Q_2+\frac{Q_1-Q_2}{\exp\left(\psi\tau \right)+1},
\end{eqnarray}
where
\begin{eqnarray}
\label{hrua}
\psi=\sqrt{\frac{\gamma}{2}}\cdot\frac{Q_1-Q_2}{T}.
\end{eqnarray}
For (\ref{29}) we claim that when the boundary conditions  satisfy the relation
\begin{eqnarray}
\label{hru2a}
\frac{\tau_R}{T}=5\sqrt{\frac{\beta\left(Q_1-Q_2\right)}{6}},
\end{eqnarray}
the solution of the equation can be expressed in terms of elementary functions:
\begin{eqnarray}
\label{oi}
Q=Q_2+\frac{Q_1-Q_2}{\left[\exp\left(\chi\tau \right)+1\right]^2},
\end{eqnarray}
where
\begin{eqnarray}
\label{chichi}
\chi=\sqrt{\frac{\beta (Q_1-Q_2)}{6}}\cdot\frac{1}{T}.
\end{eqnarray}
Note that  (\ref{oib}) doesn't goes to (\ref{oi}) when $\gamma\to 0$.  More specifically, 
when $\gamma\to 0$, the solution (\ref{oib}) goes to the weak shock solution. On the other
hand, the solution (\ref{oi}) corresponds to the case when both terms in (\ref{29}) are of the same order of magnitude.  

We used previously the linear approximation
for $V(Q)$ in the last term in the l.h.s. of (\ref{199}). Strictly speaking, since we are considering nonlinear $V(Q)$ in the r.h.s. of (\ref{199}) more consistent would be to treat  the same way the l.h.s.
Equations. (\ref{kk}), (\ref{mk}) allow us to go one step in this direction, that is  to consider
$V(Q)$ in the last term in the l.h.s. of (\ref{199}) in quadratic approximation (for cubic nonlinearity of $V(Q)$). As the result,  Eqs. (\ref{x}) and (\ref{oib}) slightly change and  Eq. (\ref{oib}) doesn't change at all.

\subsection{The shocks vs. the kinks}
\label{vs}

The main subject of the present paper is the shock waves.
However,  Eq. (\ref{199})  can describe both the shocks and the kinks \cite{kogan4}.
 To understand the difference between two types of waves note that, strictly speaking,
 the boundary conditions (\ref{bc}) tell us that the fictitious particle ends its motion at $\tau=+\infty$ at the stationary point   $Q_2$, which can be either potential minimum or potential maximum. For former case corresponds to the  shocks, the latter - to the kinks \cite{kogan4}. Equation (\ref {29}) can describe only the shocks. Equation (\ref{29b}), when the point $-Q_3$ is between $Q_2$ and $Q_1$, describes  the kink, when 
 the point is outside of that interval - the shocks.
 
Equations (\ref{x}) and (\ref{oib}) are valid both for the kinks and for the shocks, but
in these two cases the condition (\ref{x}) has different meanings. For the kinks it's the condition of their existence. For the shocks it's the condition for them to be expressed in elementary functions (actually the condition that the fictitious particle doesn't oscillate in the vicinity of the stable equilibrium). But, of course, the  shocks exist for the general boundary conditions. The same can be said about the shocks described by Eq. (\ref{29}).

\begin{acknowledgments}
We are  grateful to A. M. Kamchatnov for sending us the manuscript of his (to be published) excellent book "Theory of Nonlinear Waves",  which motivated the continuation of our previous studies.

We are very grateful to J. Cuevas-Maraver, M. Goldstein, A. Kamchatnov, E. Konig, B. Malomed, A. Novic-Cohen, V. Verbuk and A. Volkov for the insightful discussions.

We are also grateful to the organizers  of Nor-Amberd School in Theoretical Physics, the  scientific   atmosphere of which stimulated the presented research.

The paper was finalized during the Author's participation in School and Workshop on Dynamical Systems and Workshop on Localization and Ergodicity, ICTP, Trieste. The Author is grateful to the Center for the hospitality.

\end{acknowledgments}

\begin{appendix}

\section{Shock profile for the fully nonlinear transmission line}
\label{full}

Let us return to (\ref{e}).
The equation  is simplified when there is a leading mechanism of dissipation, in which cases we should substitute into (\ref{e}) either $R_C=0$ or $R_L=\infty$.
In the first case we equate the r.h.s. of (\ref{ea}) to zero and  obtain
\begin{eqnarray}
\label{v}
V=U\Phi+\text{const}_2.
\end{eqnarray}
Substituting $V$ into (\ref{qq}), solving the equation with respect to $Q$   and then substituting
 thus obtained $Q(\Phi)$ into (\ref{eb}) we obtain the equation which can be easily  integrated in quadratures.
In the second case we equate the r.h.s. of (\ref{eb}) to zero and obtain
\begin{eqnarray}
\label{qv}
I=UQ+\text{const}_1.
\end{eqnarray}
Substituting $I$ into (\ref{qq2}), solving the equation with respect to $\Phi$   and then substituting
 thus obtained $\Phi(Q)$  into (\ref{ea}) we obtain the equation which can be   integrated in quadratures.
 Thus we obtained
 Eqs.  (48a) and (48b) of Ref. \cite{kogan3}.

If both $R_C>0$ and $R_L<\infty$,  from (\ref{e}) we may obtain the equation
\begin{eqnarray}
\label{dd}
R_LR_C\frac{d Q}{d \Phi}=
\frac{U\Phi-V+\text{const}_2}{UQ-I+\text{const}_1}.
\end{eqnarray}
If we manage to solve  (\ref{dd}), (\ref{q})
we can substitute the solution into one of the equations (\ref{e}) and integrate the resulting equation in quadratures to obtain the profile of the shock wave.
It is reassuring to realize that the conditions for the integrability in elementary functions of Eqs.  (\ref{29}) and   (\ref{29b}) describing the shocks in half-nonlinear transmission line (i.e. Eqs. (\ref{x}) and (\ref{hru2a})) are actually the (nonlinear) conditions on the product $R_LR_C$. Further analysis of (\ref{dd}) we postpone until later.

\section{Solitary wave solution of a generalized KdV equation}
\label{dra}

Consider a generalized KdV equation \cite{drazin}
\begin{eqnarray}
\label{equ}
\phi_t+(n+1)(n+2)\phi^n\phi_x+\phi_{xxx}=0.
\end{eqnarray}
For a travelling wave solution
\begin{eqnarray}
\phi(t,x)=\phi(\tau),\hskip 1cm \tau=t-x/U,
\end{eqnarray}
the equation in partial derivatives (\ref{equ}) is reduced to the ODE
\begin{eqnarray}
\label{equ2}
U^3\frac{d\phi}{d\tau}-U^2(n+1)(n+2)\phi^n\frac{d\phi}{d\tau}
-\frac{d^3\phi}{d\tau^3}=0.
\end{eqnarray}

Equation (\ref{equ2})  doesn't contain explicitly the independent variable $\tau$. This prompts the idea to
consider $\phi$ as the new independent variable and
\begin{eqnarray}
\label{eee3}
E=\frac{d\phi}{d\tau}
\end{eqnarray}
as the new dependent variable.
As the result, we can present the nonlinear third order ODE (\ref{equ2}) as the linear second order ODE for $E^2$
\begin{eqnarray}
\label{difu}
\frac{d^2E^2}{d\phi^2}=2U^3-2(n+1)(n+2)U^2\phi^n.
\end{eqnarray}

A solitary wave corresponds to the boundary conditions
\begin{eqnarray}
\label{bub}
\lim_{\tau\to\pm\infty}\phi(\tau)=0,
\end{eqnarray}
which  in the new variables are \cite{kogan5}
\begin{eqnarray}
\label{boun}
E^2(0)=0, \hskip 1cm \left.\frac{dE^2}{d\phi}\right|_{\phi=0}=0.
\end{eqnarray}
Solving (\ref{difu})  with these boundary conditions we obtain
\begin{eqnarray}
\label{ps}
E^2=U^3 \phi^2-2U^2 \phi^{n+2}.
\end{eqnarray}
Substituting thus obtained $E(\phi)$ into (\ref{eee3}) and integrating we obtain the well known result \cite{drazin}
\begin{eqnarray}
\phi^n=\frac{U}{2}\sech^2\left[\frac{n\sqrt{U}}{2}\left(x-Ut\right)\right].
\end{eqnarray}

\end{appendix}

\end{document}